\documentclass[aps,prx,reprint,superscriptaddress]{revtex4-1}
\usepackage{physics}
\usepackage{graphicx}
\usepackage{upgreek}

\begin{document}


\title{A quantum light emitting diode for the standard telecom window around 1550 nm}



\author{T. M\"uller}
\email{tina.muller@crl.toshiba.co.uk}
\author{J. Skiba-Szymanska}
\affiliation{Toshiba Research Europe Ltd., 208 Science Park, Milton Road, Cambridge CB4 0GZ, UK}
\author{A. Krysa}
\affiliation{EPSRC National Epitaxy Facility, University of Sheffield, Sheffield S1 3JD, UK}
\author{J. Huwer}
\affiliation{Toshiba Research Europe Ltd., 208 Science Park, Milton Road, Cambridge CB4 0GZ, UK}
\author{M. Felle}
\affiliation{Toshiba Research Europe Ltd., 208 Science Park, Milton Road, Cambridge CB4 0GZ, UK}
\affiliation{Cambridge University Engineering Department, 9 J J Thomson Avenue, Cambridge CB3 0FA, UK}
\author{M. Anderson}
\affiliation{Toshiba Research Europe Ltd., 208 Science Park, Milton Road, Cambridge CB4 0GZ, UK}
\affiliation{Cavendish Laboratory, University of Cambridge, J J Thomson Avenue, Cambridge CB3 0HE, UK}
\author{R. M. Stevenson}
\affiliation{Toshiba Research Europe Ltd., 208 Science Park, Milton Road, Cambridge CB4 0GZ, UK}
\author{J. Heffernan}
\affiliation{Department of Electronic and Electrical Engineering, University of Sheffield, Sheffield S1 3JD, UK}
\author{D. A. Ritchie}
\affiliation{Cavendish Laboratory, University of Cambridge, J J Thomson Avenue, Cambridge CB3 0HE, UK}
\author{A. J. Shields}
\affiliation{Toshiba Research Europe Ltd., 208 Science Park, Milton Road, Cambridge CB4 0GZ, UK}

\date{\today}

\begin{abstract}
For the development of long-distance quantum networks, sources of single photons and entangled photon pairs emitting in the low-loss wavelength region around 1550 nm are a crucial building block. Here we show that quantum dot devices based on indium phosphide are capable of electrically injected single photon emission in this wavelength region with multiphoton events suppressed down to 0.11$\pm0.02$. Using the biexciton cascade mechanism, they further produce entangled photons with a fidelity of 87$\pm4$\%, sufficient for the application of one-way error correction protocols. The new material allows for entangled photon generation up to an operating temperature of 93 K, reaching a regime accessible by electric coolers. The quantum photon source can be directly integrated with existing long distance quantum communication and cryptography systems and provides a new material platform for developing future quantum network hardware.
\end{abstract}


\maketitle

\section{Introduction}
Quantum communication networks \cite{kimble_quantum_2008} are expected to enable new applications, such as cryptography secured by physical laws \cite{scarani_security_2009}, distributed quantum computing \cite{cirac_distributed_1999} and networks of geographically remote clocks \cite{komar_quantum_2014}. They require quantum infrastructure, such as quantum photon sources, non-linear quantum gates and local processing units, compatible with the low-loss fiber telecom window around 1550 nm. In particular, an essential building block for all these applications is a source of pure single photons and entangled pairs emitting in this wavelength region. A diverse range of physical systems have been put forward as candidates for these quantum hardware applications, including gallium arsenide quantum dots \cite{salter_entangled-light-emitting_2010, bennett_semiconductor_2016,de_santis_solid-state_2017}, colour centres in diamond \cite{kalb_entanglement_2017} or single atoms \cite{hacker_photonphoton_2016},  however they are all limited to wavelengths unsuitable for long distance fiber quantum network applications.\\
Quantum dots (QDs), semiconductor islands capable of confining charges in a discrete energy level structure, are particularly well-explored in the contexts of quantum communication and computing. Their implementation in InAs/GaAs, emitting around 900 nm, has provided a rich physical system to demonstrate basic building blocks of a quantum network, such as individual entangled photon pairs from electrically driven devices \cite{salter_entangled-light-emitting_2010}, photon sorters \cite{bennett_semiconductor_2016,de_santis_solid-state_2017}, and even entanglement between distant spins \cite{delteil_generation_2016,stockill_phase-tuned_2017}. Further, in contrast to downconversion sources or strongly attenuated laser pulses, their photon emission follows sub-Poissonian statistics, a pre-requisite for the most efficient quantum cryptography protocols. However, while efforts have been made to extend their emission range to longer wavelengths \cite{zinoni_c._time-resolved_2006, ward_coherent_2014,kettler_neutral_2016,paul_single-photon_2017}, electrically driven quantum light emission from QDs in the ideal telecommunication window around 1550 nm has not been possible yet.\\
On the other hand, InP-based devices such as quantum dot lasers readily reach the 1550 nm telecom window. For quantum devices, the challenge is to enable optoelectronic access to individual dots which produce light with quantum signatures, i. e. single or entangled photons. In addition, while single photon emission has been demonstrated under optical excitation from InAs dots grown by molecular beam epitaxy in the well-explored Stranski-Krastanow mode \cite{kim_d._photoluminescence_2005,cade_n._i._optical_2006,takemoto_k._optical_2007,
dalacu_deterministic_2010,benyoucef_m._telecom-wavelength_2013,dusanowski_l._single_2014,takemoto_quantum_2015}, the resulting structures are often asymmetric dashes or horns, preventing access to the low intrinsic energy splitting of the photon polarization states \cite{he_highly_2008} needed for entangled photon generation.\\
Recently, it has been shown that QD growth using metalorganic vapour phase epitaxy (MOVPE), which is the industry favoured growth method, can create InAs/InP droplet QDs with low FSS \cite{skiba-szymanska_universal_2017}. Here, we extend this growth scheme to produce the first optoelectronic devices for single and entangled photon emission in the 1550 nm telecom window.  Furthermore, we extend the working temperature up to 93 K, allowing operation with liquid nitrogen or simple closed-cycle coolers.\\

\section{Device characterisation}
\begin{figure}
 \includegraphics[width=0.5\textwidth]{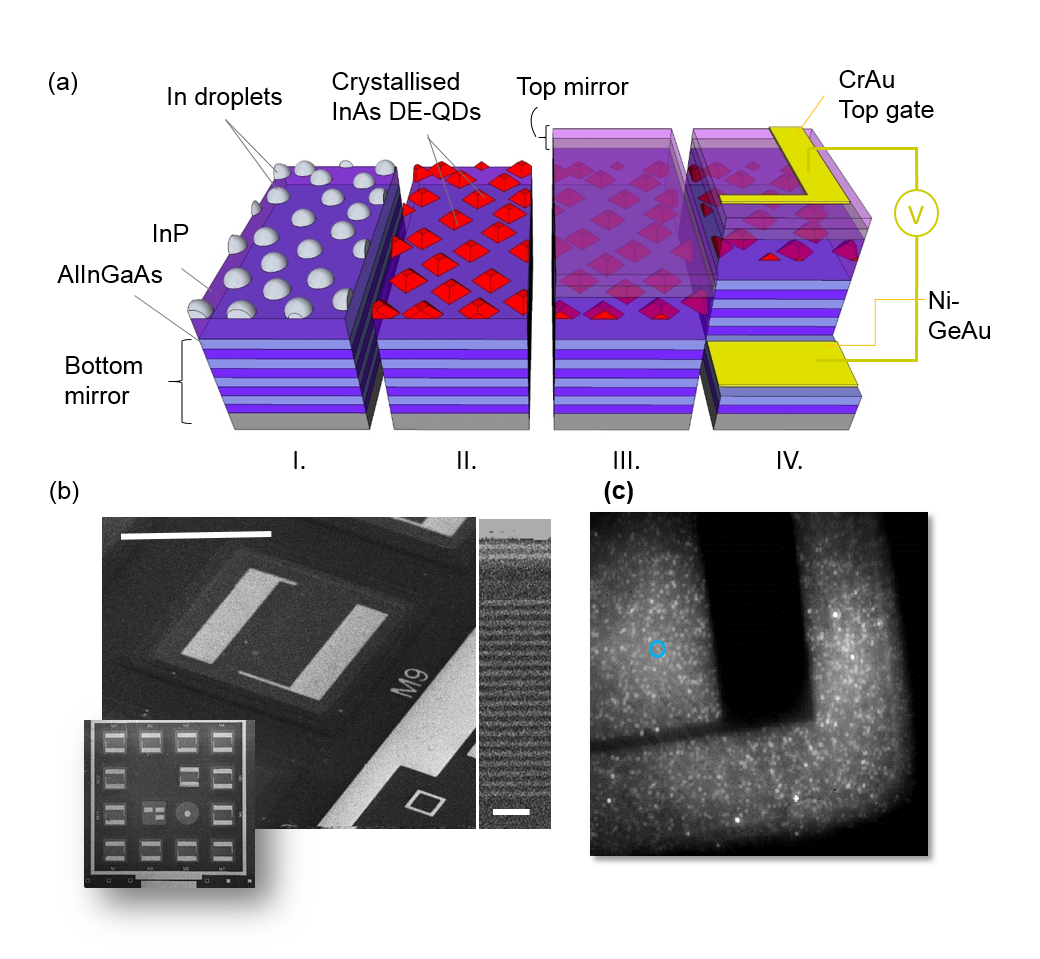}%
 \caption{Device fabrication. (a) Growth- and fabrication stages of the device, as described in the text.(b) SEM image of a typical device mesa, with a cross section through a mesa and a device overview shown in the side panel and inset, respectively. The scale bars give 500 $\upmu$m for the main panel and 500 nm for the side panel. (c)  Image of the glowing device taken with an InGaAs camera. The bright spots are individual QDs, with the dot discussed in this manuscript circled in blue.\label{Fig1}}
 \end{figure}

\begin{figure}
 \includegraphics[width=0.5\textwidth]{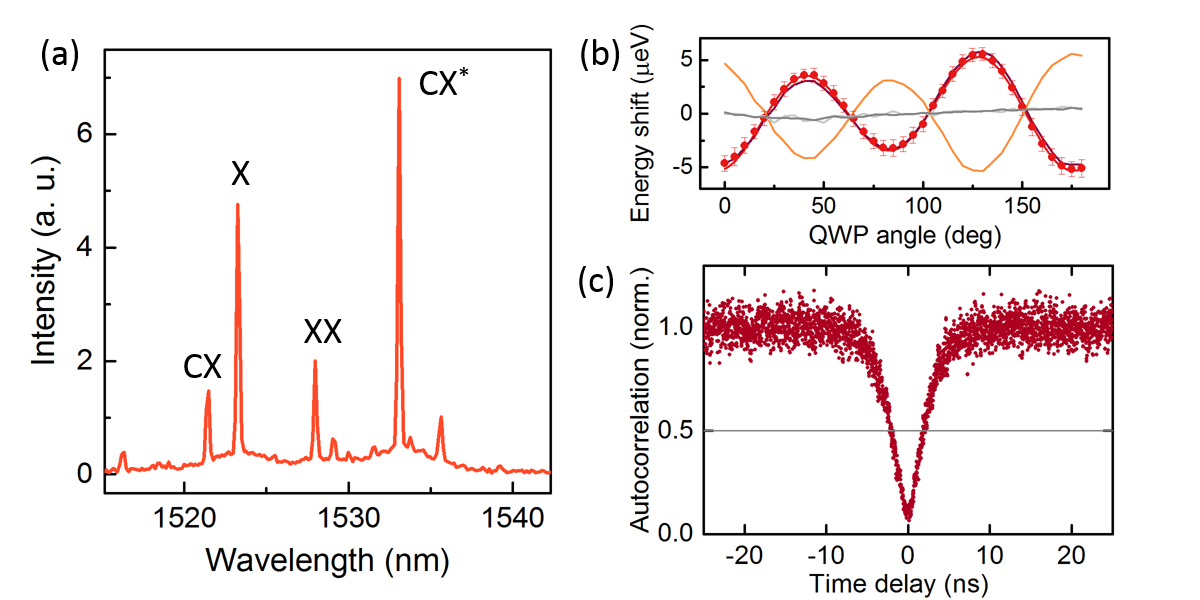}%
 \caption{Dot characterisation. (a) Spectrum of the QD circled in Fig. \ref{Fig1} (c) with exciton (X), biexciton (XX), and  charged excitons (CX and CX$^*$) labelled. The broad, low intensity features show the influence of phonon interaction at 44 K\cite{besombes_acoustic_2001}. (b) Energy shift of the X (purple curve), XX (orange curve), as well as charged CX and CX$^*$ (grey curves) transitions as a function of quarter wave plate (QWP) angle. The transition energy at each QWP angle was determined as the centre energy of a Gaussian fit to the spectrum. The solid red dots are the systematic shift corrected X data points, with the error bars giving the standard deviation derived from the fit covariance matrix. The solid red line gives the result of a fit to the QWP model described in Methods. (c) Second-order autocorrelation measurement performed on the X line in (a). \label{Fig2}}
 \end{figure}

The key features of our device are described in the cartoon in Fig. \ref{Fig1} (a).
The structure is started on the (001) surface of an InP substrate, where a 20-pair n-doped Bragg reflector is grown, each pair consisting of 112 nm of (Al$_{0.30}$Ga$_{0.70}$)$_{0.48}$In$_{0.52}$As and 123 nm of InP. After growth of a 3/4 -$\lambda$ intrinsic InP layer, metallic In is deposited for 12 s at 400 $^{\circ}$C and droplets are formed on the surface (I.). The droplets are crystallized under AsH$_{3}$ overpressure (II.) while raising the temperature up to 500 $^{\circ}$C. Next, the dots are capped with 30 nm of InP, and further overgrown with InP at 640 $^{\circ}$C. This resulted in a dot density less than 10$^9$cm$^{-2}$. After a further 5/4 -$\lambda$ intrinsic InP region, the device is finished by a three-repeat p-doped Bragg reflector to complete an asymmetric 2-$\lambda$ cavity for enhanced photon extraction (III.). For optoelectronic operation, mesas are wet-etched to the n-doped layer and the n- and p-doped layers are contacted using about 150 nm of AuGeNi slug and 20 nm of Cr followed by 100 nm of Au, respectively (IV.). A finished mesa is depicted in the SEM image in Fig. 1 (b), with a cross-section in the top panel clearly showing the layered mirrors afforded by the (001) growth surface, sandwiching the intrinsic region and QD layer. An outline of the entire device comprising 14 square mesas is shown in the inset. The resulting diode structure produces QD electroluminescence (EL) when driven in forward bias $>1.5 $ V. Using a fiber-based confocal microscope system, we can then discern individual dots as bright spots on an NIR-sensitive InGaAs camera image, as shown in Fig. 1 (c).\\
A typical EL spectrum of such a dot consists of sharp, well isolated lines, shown in Fig. \ref{Fig2} (a). The exciton (X) and biexciton (XX) transitions, upon which quantum entangled light emission is based, can be identified by monitoring their transition energy as a function of detected linear polarization: the finite exciton fine structure splitting (FSS) of 17.7$\pm$0.02 $\upmu$eV for this dot results in a variation in emission energy as shown in Fig. \ref{Fig2} (b), where the quarter-wave plate method described earlier \cite{skiba-szymanska_universal_2017} was used. In contrast, the energy of charged transitions is independent of detected polarization. Both the X and XX transitions have linewidths limited by the spectrometer resolution (89 $\mu$eV), and, after being filtered and guided to our SSPD detection system, result in count rates up to 200 kcps. Further details on the specifics of our setup and detector efficiencies are given in Methods.\\
To determine the quantum nature of the observed transitions, we performed intensity autocorrelation ($g^{(2)}$) measurements, as shown in Fig. \ref{Fig2} (c) for the exciton transition of the dot in Fig. \ref{Fig1} (c). The dip in coincidences for zero delay is well below 0.5, with the measured $g^{(2)}(0)=0.11\pm0.02$, which proves emission from a true single photon source. This value does not include corrections due to dark and background counts or detector jitter and therefore gives an upper bound to multiphoton emission from our device. Even without further optimisation, it is suppressed by almost a factor 10 compared to a Poissonian photon source.\\
\section{Entangled photon generation}
\begin{figure}[h!]
 \includegraphics[width=0.5\textwidth]{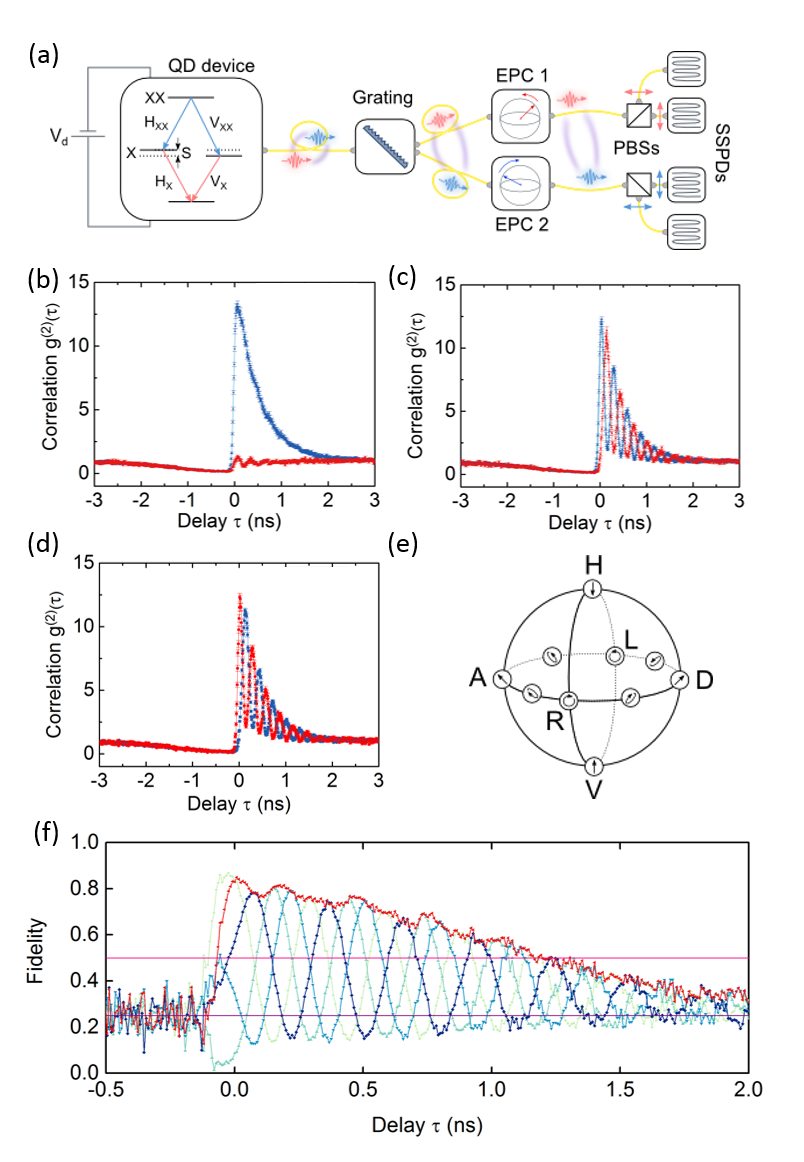}%
 \caption{Measurement of entanglement fidelity. (a) Schematic drawing of the entanglement setup. The entangled photons are generated by the biexciton cascade shown diagrammatically in the first panel. X and XX photons are separated by a diffraction grating and sent to their respective polarization sensitive detection units.  These consist of electronic polarization controllers (EPCs) and polarising beam splitters (PBSs) to prepare the detection system in one of five measurement bases. Polarization selected photons are measured using superconducting single photon detectors (SSPDs), which have efficiencies ranging from 40\% to 60\%. Photon arrival times are registered and compared using a time interval analyzer. (b), (c) and (d)  Co-polarized (blue curves) and cross-polarized (red curves) biexciton-exciton photon coincidences measured in the HV, DA and LR bases, respectively. (e)  Schematic of the Poincar\'e sphere, with measured bases indicated by cartoons of the photon polarization.  (f) Entanglement fidelity to four maximally entangled states with phases $\chi = 0$, $\frac{\pi}{2}$, $\pi$, and $\frac{3\pi}{2}$ (light to dark blue curves), as well as to an evolving state (red curve). Pink and purple lines give the classical limit and uncorrelated values for coincidences, respectively.\label{Fig3}}
 \end{figure}

Next, we show that we can generate entangled photons with our telecom-wavelength diode. We use the biexciton cascade mechanism \cite{benson_regulated_2000}, with the measurement configuration schematically shown in Fig. \ref{Fig3} (a). Starting out with the QD in the biexciton state, radiative recombination of a first electron-hole pair leaves the dot occupied by a single exciton. Due to conservation of angular momentum, the spins of the remaining charge carriers are entangled with the polarization of the emitted biexciton photon. Recombination of the exciton therefore produces a photon in a polarization state predefined by the path taken during the first recombination step, and leads to the emission of entangled photons in the maximally entangled state \cite{stevenson_evolution_2008}
\begin{equation}
\ket{\Psi_B(\tau)}=\frac{1}{\sqrt{2}}\left(\ket{H_{XX}H_X}+\exp(\frac{iS\tau}{\hbar})\ket{V_{XX}V_X}\right).
\end{equation} The phase factor here is acquired during the time $\tau$ spent in the exciton state: the fine structure splitting $S$ between the two spin states leads to a time-dependent oscillating correlation between the polarization of the two photons of the form $\cos^2 (\frac{s\tau}{2\hbar})$ \cite{ward_coherent_2014}. \\
For the dot presented in Fig. \ref{Fig2}, the XX-X polarization correlations measured in the dot eigenbasis HV are shown in Fig. \ref{Fig3} (b). For negative time delays we find the expected dip in coincidences, corresponding to the unlikely arrival of an X photon before a XX photon. For positive time delays, co-polarized coincidences at their maximum are more than ten times stronger than an uncorrelated source, and decay for increasing time delays. This can be understood as follows: subsequently emitted XX and X photons are only correlated as long as they stem from the same excitation cycle. If the time delay between the two photons becomes comparable to the natural lifetime of the X, or the timescale of any other mechanism destroying the X population such as re-excitation, the likelihood of measuring uncorrelated photons increases, until, for long delays, only uncorrelated photons remain. In contrast, cross-polarized coincidences never significantly exceed the Poissonian value. The effect of the finite $S$ in this dot is evident in the polarization coincidences measured in a superposition basis, as shown in Fig. \ref{Fig3} (c) for the example of the diagonal/antidiagonal (DA) basis and in Fig. \ref{Fig3} (d) for the left-hand circular/right-hand circular (LR) basis. For positive delays, the coincidences follow the oscillatory pattern outlined above. Similar to the HV case, the envelope is limited by the X radiative lifetime and re-excitation. \\
To prove that emission from our device is entangled and determine the fidelity to an evolving Bell state \cite{ward_coherent_2014,stevenson_evolution_2008}, we performed the same coincidence measurements in two additional bases: the two elliptical bases (E$_{LD}$E$_{RA}$ and E$_{LA}$E$_{RD}$) as marked on the Poincar\'e sphere in Fig. \ref{Fig3} (e). For unpolarized dot emission, the degree of correlation $C_{mn}$ between the polarization basis states m and n is derived by dividing the difference between orthogonal coincidences by their sum, which gives the contrast between co- and cross-polarized coincidences. The fidelity to the expected ideal evolving state is then given by \cite{ward_coherent_2014}
\begin{eqnarray}
F(t)&=&\frac{1}{4} \left[1+C_{HV}+(C_{DA}-C_{LR} )\cos(\frac{S\tau}{\hbar})\right. \nonumber\\
& &\left.+(C_{E_{LD} E_{RA}} -C_{E_{LA} E_{RD}})  \sin(\frac{S\tau}{\hbar}) \right],
\end{eqnarray}
which also expresses the fidelity to static Bell states for fixed values of  $\chi=\frac{S\tau}{\hbar}$. The experimental results are shown in Fig. \ref{Fig3} (f) for both the evolving and the static Bell states. The fidelities of our two-photon emission to all states clearly violate the limit of 0.5 attainable by a classical photon source. The maximum fidelity of 0.87$\pm$ 0.04, more than 9 standard deviations above the classical limit, is obtained for $\chi=0$, corresponding to the symmetric Bell state. For the evolving choice of basis, it is straightforward to see that entanglement persists for over a ns, eliminating the need for high-resolution temporal post-selection. This even holds for static measurement bases, where the effect of the FSS can be effectively erased by employing appropriate measurement techniques \cite{wang_-demand_2010, fognini_universal_2017}. We also evaluated the corresponding Bell's parameters \cite{clauser_proposed_1969}, where the classical limit of 2 was violated by all states and a maximum of 2.35$\pm$0.27 was reached. For long time delays, the fidelity drops to the uncorrelated value of 0.25, once exciton emission from separate excitation cycles begins to dominate.\\
For quantum communication applications, error correction protocols can compensate for the effects of non-perfect entangled photon sources. For all bases, the fidelity reaches above the threshold of 70\%, corresponding to a quantum bit error rate of 20\%, required for two-way protocols \cite{chau_practical_2002}. The entangled light source is therefore suitable for quantum communication protocols enabling secure data networks protected by the laws of physics. In addition, for one-way error correction protocols such as CASCADE to apply, the quantum bit error rate has to be below 11\% \cite{brassard_g._advances_1994}, which corresponds to an entanglement fidelity of 83.5\%. This limit is surpassed by both the symmetric and the evolving bases.  \\
\section{Entangled photons at elevated temperatures}
\begin{figure}
 \includegraphics[width=0.5\textwidth]{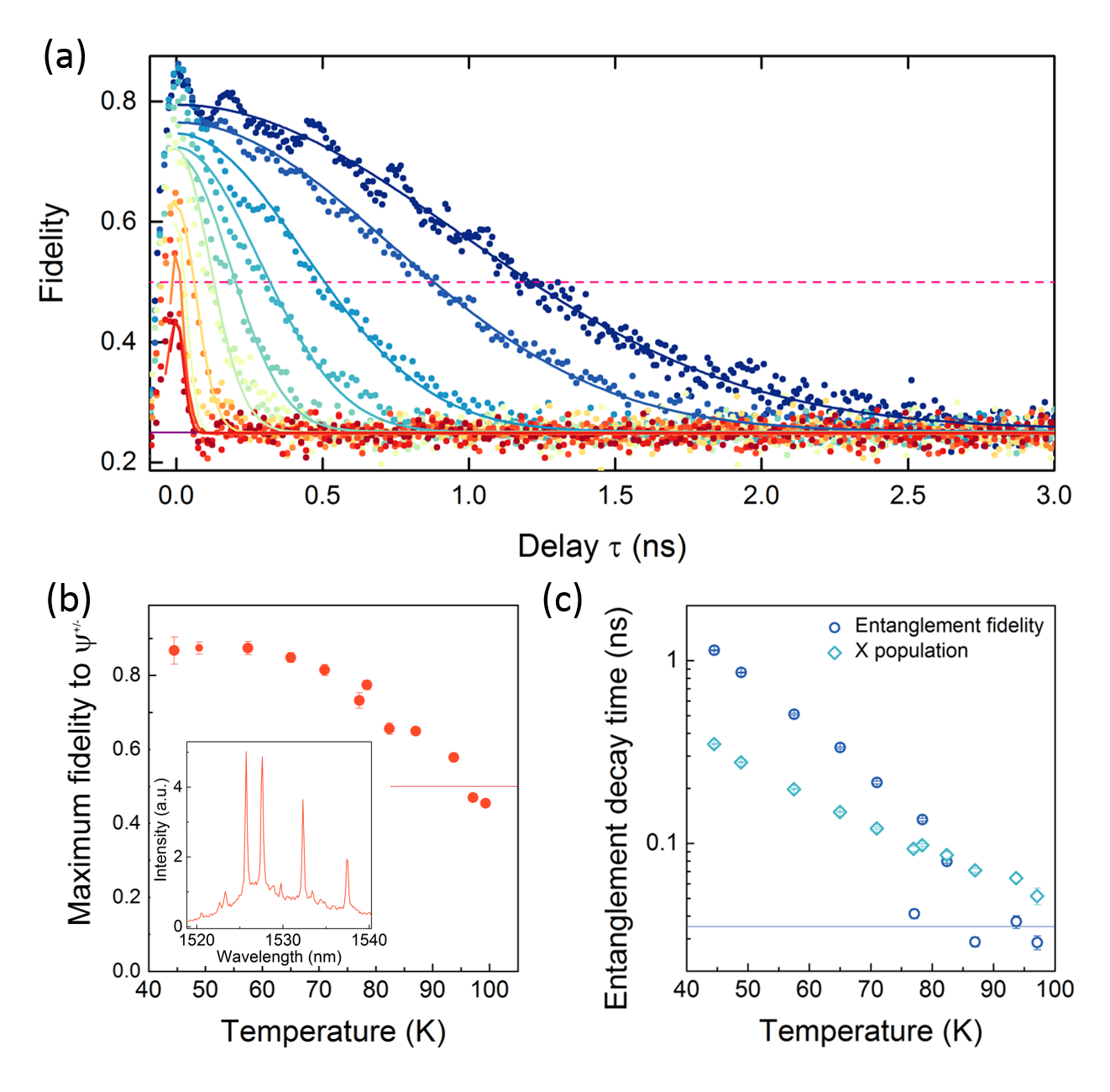}%
 \caption{Entanglement at elevated temperatures. (a) Fidelity to an evolving maximally entangled state when increasing the sample temperature from 44 K (dark blue) to 99 K (dark red) for a constant driving voltage of 1.8 V. Measured data (solid dots) are fitted with a Gaussian decay (solid curves), and the classical threshold of 0.5 is indicated by the dashed pink line. (b) Maximum fidelity to the $\Psi^+$ state as a function of temperature, where the error bars are deduced from Poissonian statistics on the correlations. Again, the classical limit is indicated by the solid red line. Inset: dot spectrum at 63 K. (c) Entanglement decay time constant extracted from the Gaussian fits in (a) (blue circles), and X decay time (turquoise diamonds,) with the error bars giving the standard deviation calculated from the fit covariance matrix. The detector timing resolution is around 34 ps, as shown by the solid blue line.\label{Fig4}}
 \end{figure}

Finally, we show that droplet epitaxy QDs based on InAs/InP are able to produce entangled photons even at elevated temperatures. Fig. \ref{Fig4}(a) shows the measured fidelity to an evolving Bell state while increasing the temperature from 44 K to 99 K, for a constant driving voltage of 1.8 V. The extracted maximum fidelities for each temperature are given in Fig. \ref{Fig4} (b), and they remain above the classical limit of 0.5 up to a temperature of 93.6 K. This indicates that the spin relaxation quenching observed in GaAs QDs  \cite{paillard_spin_2001} does not affect InP based dots as strongly, and consequently the operating temperature of our entangled photon diode exceeds those obtained for GaAs devices  \cite{dousse_quantum_2010}.\\
Two independent effects are limiting the entanglement fidelity for elevated temperatures in our system. For temperatures below around 78 K, phonon broadening \cite{besombes_acoustic_2001} causes photons from transitions other than X and XX to be emitted at these transition wavelengths, leading to uncorrelated detection events. As an example, the spectral broadening at 63 K is shown in the inset to Fig. \ref{Fig4} (b).  Increasing the temperature also leads to faster decay timescales of entanglement fidelity, as seen in Fig. \ref{Fig4} (a), and quantified by the Gaussian HWHM plotted in Fig. \ref{Fig4} (c).  This is due to the combined reduction of the timescales involved in the biexciton dynamics, which is shown for the example of the X decay time measured from unpolarised XX-X crosscorellations in Fig. \ref{Fig4} (c). Note that the entanglement fidelity persists for longer than the exciton lifetime for all temperatures below 78 K. This is an indication that uncorrelated events start to dominate only at timescales longer than the X decay time and that other processes disruptive of X population, such as re-excitation to higher levels, therefore must be present. Around 78 K, the fidelity decay time becomes comparable to the $<$ 34-ps timing resolution of our detectors [blue solid line in Fig. \ref{Fig4} (c)]. The convolution with the Gaussian detector response then leads to an effective averaging and decrease of the maximum fidelity beyond the phonon broadening effect. Higher resolution detector electronics and pulsed operation, together with a diode structure optimized for a reduction in CX intensity, would therefore increase the operation temperature of an entangled photon LED even further. \\
\section{Conclusion}
The InAs/InP droplet epitaxy quantum dot devices we introduce here are expected to have significant impact in the development of quantum network technology. Unlike their InAs/GaAs predecessors, their emission wavelength near 1550 nm will crucially enable them to interface with established quantum key distribution technology, and extend the reach of quantum networks with quantum relays and repeaters. Superior performance is anticipated compared to light sources based on optical nonlinearities, by suppression of intrinsic errors due to sub-Poissonian character of the emitted light. Pulsed operation of smaller versions of the devices is immediately feasible once integrated with suitable RF electronics, since the physical mechanisms remain the same. Further, their compatibility with standard industry fabrication techniques, use of materials dominant in 1550 nm photon detectors, and amenability to miniaturisation and on-chip integration afforded by electrical operation makes them an attractive and practical building block for integrated quantum network transceivers, and engineering of the quantum internet.

\begin{acknowledgments}
We thank Marco Lucamarini, Anthony Bennett and Peter Spencer for discussions. This work has been co-funded by the UK's innovation agency, Innovate UK, and the EPSRC.
\end{acknowledgments}


%

\appendix*
\section{Methods}
\subsection*{Growth and fabrication of devices}
On an InP substrate, we use MOVPE to grow 20 pairs of DBR, each pair consisting of 112 nm of (Al$_{0.30}$Ga$_{0.70}$)$_{0.48}$In$_{0.52}$As and 123 nm of InP, both n-doped with Si (doping concentration $2\times10^{18}$ cm$^{-3}$). After a 3/4 lambda intrinsic InP layer, formation of In droplets was achieved by supply of In to the InP surface for 12 s, via pyrolysis of trimethylindium at 400 $^o$C while withholding the supply of arsine to the growth chamber. The QD crystallization process under arsine overpressure started at 400 $^o$C and carried on until the substrate reached 500 $^o$C. Next the dots were capped with 30nm of InP followed by more InP at 640 $^o$C. This resulted in a dot density less than 10$^9$ cm${-2}$. A further 5/4 lambda intrinsic InP was grown to cover the QDs, before the cavity was finished off with three repeats of p-doped top DBR (doping concentration $2\times10^{18}$ cm$^{-3}$). Note that although the nominal dot composition is pure InAs, we can assume it is highly likely to be InAsP in reality, as the optical signal of a 2D layer formed by As/P interchange during dot cystalisation is observered. To contact to the n- doped layer, about 150 nm of AuGeNi slug was deposited and annealed at 375  $^o$C. The p-doped layer was contacted by depositing around 20 nm of Cr, followed by 100 nm of Au. A detailed characterisation of a similar dot growth protocol for undoped structures was reported earlier \cite{skiba-szymanska_universal_2017}.
\subsection*{Fiber-based confocal microscope}
The device is placed in a temperature controlled environment in a He vapour cryostat, and navigation to a desired dot is achieved through an imaging system using an InGaAs camera which is highly sensitive in the NIR region and piezo-driven xyz stages. Photons are collected using a confocal microscope with a high-NA lens (NA$=0.68$) and coupled to a single-mode fibre which acts as a pinhole. The light is then guided either to a spectrometer or to a free-space grating setup to single out the biexction and exciton photons with a wavelength resolution $\sim$0.5 nm. The light is then fibre-coupled again and sent to electrical polarisation controllers to prepare the light in one of the five basis states, before being polarisation filtered at a polarising beam splitter and sent to superconducting single photon detectors with efficiencies ranging from 45\% to 60\% and timing jitters between 60 – 68 ps as determined from the full width at half maximum (FWHM) of a Gaussian fit. Photon arrival times are registered and compared using a time interval analyser (smallest time bin 32 ps).
\subsection*{QD transition identification}
Neutral excitons and biexcitons are easily assigned a transition through the FSS measurement described in the main text. Positively and negatively charged excitons are identified by the behaviour at higher temperatures, where the positively charged exciton becomes more prominent as the limiting holes are supplied at greater abundance for elevated temperatures. Other less intense lines are due to further charged transitions as well as other dots nearby. 
\subsection*{FSS measurement}
Our method uses a quarter wave plate and subsequent polariser directly in front of the spectrometer, as opposed to the more commonly used half wave plate/polariser combination. It has the advantage that elliptical polarisations can be detected, which allows us to correct for birefringence in our setup optics.  As described earlier\cite{skiba-szymanska_universal_2017}, the measured energy deviation $\Delta E$ from the mean energy $\epsilon$ of a state with polarisation $p$ (typically zero) as a function of the quarter wave plate angle $\chi$, with fitting parameters $\theta$ and $\phi$ describing the polarisation rotation and acquired phase shift caused by the setup optics, respectively is given by
\begin{widetext}
\begin{eqnarray*}
   \Delta E (\chi)&=& E(\chi)-\epsilon\\
&=&  \frac{s}{2}\left(\frac{2p+\cos\theta(1+\cos4\chi)+\sin\theta \sin4\chi \cos\phi-2 \sin\theta \sin2\chi \sin\phi}{2+p \cos\theta (1+\cos4\chi)+p \sin\theta \sin4\chi \cos\phi-2p \sin\theta \sin2\chi \sin\phi }\right). 
\end{eqnarray*}	
\end{widetext}
\subsection*{Calibration of the polarisation states} For the calibration of the electronic polarisation controllers, a polarised light source was used to mimic QD emission. Its polarisation was aligned with the QD axes for the calibration of the HV basis, and then set to the desired basis using a half wave plate followed by a quarter wave plate. The electronic polarisation controllers were independently varied in each basis to minimise transmission through the polarising beam splitters, effectively mapping the desired basis to the HV system. Co- and cross-polarised measurements were then taken using the transmission and reflection arms of the fibre based beam splitter. In practice, small deviations in the EPC alignment from the QD bases can occur, which are responsible for small admixtures of photons measured in bases other than the intended. This is the reason that remaining signatures of evolving states (oscillations) are observed in the HV correlations in Fig. \ref{Fig2} (c).


\end{document}